\documentclass[a4paper,11pt]{article}
\pdfoutput=1 % if your are submitting a pdflatex (i.e. if you have
\usepackage{jheppub} % for details on the use of the package, please
\usepackage{natbib}
\renewenvironment{description}[1][0pt]
{\list{}{\labelwidth=0pt \leftmargin=#1
		}}
{\endlist}                                        
\usepackage{tensor}
\renewenvironment{description}[1][0pt]
{\list{}{\labelwidth=0pt \leftmargin=#1
		}}
{\endlist}    
\usepackage[T1]{fontenc} % if needed

\title{\boldmath Is Teleparallel Gravity really equivalent to General Relativity?}

\author[a]{Luciano Combi,}
\author[a,b]{Gustavo E. Romero}
\affiliation[a]{Instituto Argentino de Radioastronom\'ia (CCT-La Plata, CONICET; CICPBA), C.C. No. 5, 1894, Villa Elisa, Argentina}
\affiliation[b]{Facultad de Ciencias Astron\'omicas y Geof\'isicas, Universidad Nacional de La Plata, Paseo del Bosque s/n, 1900 La Plata, Buenos Aires, Argentina.}
\emailAdd{lcomb@iar.unlp.edu.ar}
\abstract{An axiomatization of the so-called Teleparallel Equivalent
to General Relativity is presented. A set of
formal and semantic postulates are elaborated from where the physical meaning
of various key concepts of the theory are clarified. These concepts include those of
inertia, Lorentz and diffeomorphism invariance, and 
reference frame. It is shown that
Teleparallel Gravity admits a wider representation of
space-time than General Relativty, allowing to define
properties of the gravitational field such as energy and
momentum that are usually considered problematic.
In this sense, although the dynamical equations of both
theories are equivalent, their inequivalence from a physical
point of view is demonstrated. Finally, the axiomatic
formulation is used to compare Teleparallel Gravity with
other theories of gravity based on absolute parallelism
such as non-local and f(T) gravity.}

\begin{document} 
\maketitle
\flushbottom

\section{\label{sec:level1} Introduction}

In 1928, Einstein attempted to formulate a unified theory of gravity and electromagnetism \cite{Einstein1930} using the geometrical notion of teleparallelism (called \textit{Fernaparallelismus}, in German), a concept developed independently by Cartan a few years before. In this new theory, the metric is replaced by the tetrad field $e^{a}_{\mu}$, a 16-component object which would encode the 10 degrees of freedom of the metric and the 6 degrees of freedom of the electromagnetic field. Even though Einstein was not able to find consistent fields equation for his theory, the idea of an alternative representation for the gravitational interaction using torsion instead of curvature was introduced. This theory, with field equations equivalent to those of General Relativity (GR) and a different geometrical background, is now called the Teleparallel Equivalent to General Relativity or the Teleparallel Framework of General Relativity (GR$_{||}$). 

GR$_{||}$ admits a well-behaved (though gauge dependent) enhhergy-momentum tensor for gravity \cite{moller1961tetrad}. Adopting the usual interpretation of tetrad fields as reference frames, Maluf and collaborators have shown that the concept of gravitational energy in GR$_{||}$ is consistent in many physical situations (see Ref. \cite{Maluf:2013gaa}, \cite{Ulhoa:2010wv} and \cite{Maluf:2007qq}). Moreover, this approach has been used by Mashhoon to formulate a non-local gravity theory \cite{Mashhoon:2014twa}. Despite all these interesting developments, a thorough comparison of GR and GR$_{||}$ has not been made. In this paper, we have constructed an axiomatization for GR$_{||}$ with the aim of implementing a rigorous comparison between both theories. We have analyzed and compared all key physical concepts in GR and GR$_{||}$ in a rigorous and systematic way. We have then analyzed the physical interpretation of alternative teleparallel theories, such as $f(\mathbb{T})$ gravity, that are currently under investigation. We have restricted ourselves to the so-called pure tetrad approach to GR$_{||}$ where we assume a preferred frame to construct the teleparallel geometry. However, there is an alternative approach based in the translation gauge interpretation \cite{Aldrovandi:2013wha}, where the local Lorentz invariance in maintained. A thorough discussion of the differences between both approaches is outside the scope of this paper and will be treated elsewhere

Our paper is organized as follows. Firstly we present GR$_{||}$ in a heuristic way in Section 2. Then we proceed to develop the axiomatization of the theory in Section 3. Section 4 includes a detailed comparison of both theories. We also include an exhaustive characterization of the teleparallel energy-momentum tensor and a comparison between GR$_{||}$ and other teleparallel theories. Finally, we present our conclusions in Section 5.

\section{Teleparallel Gravity}

We begin this section revising the geometrical framework of GR$_{||}$: the affine Weitzenb\"ock geometry. Then, we move to describe the main features of the field formulation of the theory. From all these elements, we will construct a rigorous axiomatization.

\subsection{Geometrical background}
The dynamical object of GR$_{||}$ is the tetrad, or \textit{vierbein}. Given a pseudo-Riemannian manifold $\mathcal{M}$, a tetrad is an orthonormal basis field $\bold{e}_{a}= e_{a}^{\mu} \partial_{\mu}$ ($a=0,..,3$) of the tangent bundle $T\mathcal{M}$. The co-frame is denoted as $\bold{e}^{a}$, holding
\begin{equation}
e^{a}_{\mu} e_{b}^{\mu} = \delta^{a}_{b}.
\end{equation} 

The tetrad encodes the metric structure of the manifold as
\begin{equation}
g_{\mu \nu}=e^{a}_{\mu}e^{b}_{\nu} \eta_{ab},
\end{equation}
where $\eta^{ab}=\text{diag} (-1,1,1,1)$ is the Minkowski metric in Cartesian coordinates. We adopt the distinction between Greek letters $\mu,\nu,..$ for space-time coordinate indices, and Latin letters $a,b,..$ for Lorentzian tangent-space indices. 

If the manifold admits a global smooth frame, then it is called \textit{parallelizable}. The transformation group of the tetrad that preserves orthonormality is SO$(1,3)$. In general, these transformations $\Lambda^{a}_{b'}(x)$ are point-dependent:
\begin{equation}
\mathbf{e}_{b'} (x)= \Lambda^{\; \; a}_{b'} (x) \mathbf{e}_{a} (x).
\end{equation}

A useful characterization of the tetrad field is given by the Lie bracket:
\begin{equation}
[\mathbf{e}_a,\mathbf{e}_{b}]= \Omega^{c}_{ab} \mathbf{e}_c,
\label{eq: coef}
\end{equation} 	
where $ \Omega^{c}_{ab} \equiv e^{\mu}_{a} e_{b}^{\nu} \partial_{[\nu} e^{c}_{\mu]} $ are the structural or anholonomy coefficients. The importance of these coefficients lays on the Frobenius theorem \cite{Fecko:2006zy}: iff these coefficients are null for a smooth tetrad $e^{a}_{\mu}$, there exists a coordinate system $\lbrace x^{\mu} \rbrace$ such that
\begin{equation}
e^{a}_{\mu} = \delta^{a}_{\mu}.
\end{equation}

In order to formulate the dynamics of the theory we need to introduce an affine structure in the manifold. For instance, GR is formulated with the only connection that is uniquely fix by the metric, the Levi-Civita connection:
\begin{equation}
\Gamma^{\rho}_{\mu \nu}:= \frac{1}{2} g^{\rho \sigma} \Big ( \partial_{\mu} g_{\nu \sigma} + \partial_{\nu} g_{\sigma}-\partial_{\sigma} g_{\mu\nu} \Big ).
\label{eq: christoffel}
\end{equation}

As it is well known, this connection is metric-compatible and torsion-free. In this way, the field equations for GR are constructed from the Riemann curvature, which represents the presence of gravitational interaction. On the other hand, given a smooth tetrad field $\bold{e}_{a}$, we can induce on the manifold the Weitzenb\"ock connection:
\begin{equation}
^{*} \Gamma^{\rho}_{\mu \nu}:= e^{\rho}_{a} \partial_{\nu} e^{a}_{\mu}.
\label{eq: Wei}
\end{equation}

It can be checked that (\ref{eq: Wei}) is a metric-compatible and curvature-free connection. The absence of curvature indicates the presence of absolute parallelism on the manifold; this means that two vectors in different tangent spaces are parallel if their projections on the tetrad are proportional, regardless the path connecting both spaces. The covariant derivative of a vector can be written as:
\begin{equation}
^{*} \nabla_{\nu} V^{\lambda} := \partial_{\nu} V^{\lambda} + ^{*} \Gamma ^{\lambda}_{\mu \nu} V^{\mu} = e^{\lambda}_{a} \partial_{\nu} ( e^{a}_{\mu} V^{\mu} ) = e^{\lambda}_{a} \partial_{\nu} V^{a},
\label{eq: covvec}
\end{equation}
i.e, $V^{\lambda}$ is parallel transported if $V^{a}$ is constant \cite{Ferraro:2011us}. From \ref{eq: covvec} we also get that the tetrad is trivially parallel-transported:
\begin{equation}
\; ^{*} \nabla_{\nu} e^{\lambda}_{a} = 0.
\label{eq: para}
\end{equation}

The fundamental tensor of the Weitzenb\"ock geometry is the torsion tensor:
\begin{equation}
T^{\rho}_{\; \; \mu \nu}:=  2 \; ^{*}\Gamma^{\lambda}_{[ \mu \nu ]} \equiv e^{\rho}_a \Big ( \partial_{\nu} e^a_{\mu} - \partial_{\mu} e^a_{\nu} \Big )=g^{\rho \sigma}  T_{\sigma \mu \nu},
\end{equation}
which is antisymmetric in the last two indices. The Weiztenb\"ock connection is linked with the usual metric Levi-Civita connection by the so-called \textit{contorsion} tensor
\begin{equation}
^{*} \Gamma ^{\rho}_{\mu \nu} = \Gamma^{\rho}_{\mu \nu} + K^{\rho}_{\mu \nu},
\end{equation}
related to the torsion tensor as:
\begin{equation}
K_{\; \mu \nu}^{\rho}=\frac{1}{2} g^{\rho \sigma} \Big ( T_{\nu \sigma \mu} + T_{\mu \sigma \nu} - T_{\sigma \mu \nu} \Big )= g^{\rho \sigma} K_{\sigma \mu \nu},
\label{eq: contortion}
\end{equation}
and then
\begin{equation}
U^{\mu} \nabla_{\mu} U^{\rho}=0 \rightarrow U^{\mu} \; ^{*}\nabla_{\mu} U^{\rho} = K^{\rho}_{\sigma \nu} U^{\sigma} U^{\nu}.
\end{equation}

Let us note that the affine geometry induced on the manifold is an independent concept of the metric structure (a fact first noted by Cartan \cite{Sauer2006399}). In other words, given a metric field, we can build several connections over it. This is a relevant issue for the proper representation of what we call space-time (see Section 4). 

The teleparallel condition, encoded in \ref{eq: para}, is established for a certain preferred tetrad $\mathbf{e}_{a}(p)$. If we perform a local Lorentz transformation $\Lambda^{a}_{b'}(p)$ of the frame, we preserve the metric structure but the Weitzenb\"ock parallelism is broken \cite{local}. This leads to non covariant transformations of all Weitzenb\"ock tensors such as torsion:

\begin{equation}
T^{\mu}_{\nu \rho} [ \bold{e}_{a}]= T^{\mu}_{\nu \rho} [ \bold{e}_{b'} ] + e^{\mu}_a e^{b}_{\rho} \omega^{a}_{\nu b} - e^{b}_{\nu} e^{\mu}_{a} \omega^{a}_{\rho b},
\end{equation}
where $\omega^{a}_{\nu b}:= \Lambda^{a}_{c} \partial_{\nu} \Lambda^{c}_{b}$ is known as the flat Lorentz connection. The invariance of the Weitzenb\"ock geometry under global Lorentz transformation ---but not under local ones--- entails a precise physical meaning once we assume that the tetrad frame represents a reference system (see Section 4.2). In the following, we review how to achieve a geometrized theory of gravity based on a teleparallel geometry.

\subsection{Dynamics}

Analogous to curvature in the Riemannian geometry of GR, in teleparallel theories, the Weitzenb\"ock torsion is used to build the Lagrangian and to obtain the field equations for the tetrad frame. Let us begin with a general Lagrangian constructed with quadratic Weitzenb\"ock scalars:

\begin{equation}
\mathcal{L} = \kappa \Big ( a_1 T^{\rho}_{\; \; \mu \nu} T^{\; \;  \mu \nu}_{\rho} +a_2 T^{\rho}_{\; \; \: \mu \nu} \tensor{T}{^\mu^\nu_\rho} +a_3 T^{\rho}_{\; \; \mu \rho} \tensor{T}{^\nu^\mu_\nu} \Big ).
\end{equation}
\textbf{being $T^{\mu}_{\; \sigma \rho} = g_{\nu \sigma} \tensor{T}{^\mu^\nu_\rho}= g^{\mu \alpha} g_{\beta \sigma} g_{\nu \rho} T^{\; \;  \mu \nu}_{\alpha}$}. This general teleparallel Lagrangian was first studied by Hayashi et al. \cite{Hayashi:1979qx}, who explored a set of theories called New General Relativity. If we choose the constants as $a_1=1/4$, $a_2= 1/2$ and $a_3 =-1$, we obtain the Lagrangian $\mathcal{L}_{T}$ of GR$_{||}$. This Lagrangian can be casted as:
\begin{equation}
\mathcal{L}_{T} \equiv \kappa \mathbb{T} =  \kappa \Sigma^{\rho \mu \nu} T_{ \rho \mu \nu},
\end{equation}
where $\mathbb{T}:=\frac{1}{4} T^{\rho}_{\; \; \mu \nu} T^{\; \;  \mu \nu}_{\rho} +\frac{1}{2} T^{\rho}_{\; \; \mu \nu} \tensor{T}{^\mu^\nu_\rho} - T^{\rho}_{\; \; \mu \rho} \tensor{T}{^\nu^\mu_\nu} $ is the \textit{torsion} scalar and
\begin{equation}
\Sigma^{\rho \mu \nu}:= \frac{1}{2} \Big ( K^{\mu \nu \rho} -g^{\rho \nu} T^{\sigma \mu}_{\; \; \; \sigma} + g^{\rho \mu} T^{\sigma \nu}_{\; \; \; \sigma} \Big ),
\end{equation}
is the \textit{superpotential}. This particular choice of coefficients allows the following noteworthy decomposition\cite{local}:
\begin{equation}
\mathbb{T} \equiv  -R  -  2 \nabla^{\mu}  T^{\nu}_{\; \; \mu \nu},
\label{eq: riccitorsion}
\end{equation}
where $R$ is the Riemannian Ricci scalar. Since both scalars differ by a boundary term, the dynamic equations of GR$_{||}$ are equivalent to Einstein equations. The complete action of GR$_{||}$ is then:
\begin{equation}
S[\mathbf{e}^{a}]= -\kappa \int  \mathbb{T} \: \sqrt{-g}  \: d^4x 	+ S_{M}.
\end{equation}

Whereas the Ricci scalar has second-order derivatives on the metric, the teleparallel Lagrangian has first derivatives on the tetrad field. This allows to use the Euler-Lagrange equations:

\begin{equation}
\frac{\partial \mathcal{L}}{\partial e_{a \rho}} -\partial_{\sigma} \frac{\partial \mathcal{L}}{\partial(\partial_{\sigma} e_{a \rho} )} = \frac{\delta \mathcal{L}_M}{\delta e_{a \rho}},
\end{equation}
from which we get the Yang-Mill type equations
\begin{equation}
\partial_{\nu} \Big ( \sqrt{-g} \: \Sigma^{a \lambda \nu} \Big ) = \frac{\sqrt{-g}}{4k} \Big ( t^{\lambda a} + \Theta^{\lambda a}  \Big ),
\label{eq: campos}
\end{equation}
where we use
\begin{equation}
\frac{\partial \mathcal{L}	}{\partial( \partial_{\sigma} e_{a \lambda})}= - 4 \kappa \sqrt{-g} \: \Sigma^{a \lambda \sigma},
\end{equation}
\begin{equation}
	 - \sqrt{-g} \: t^{\lambda a}:=\frac{\partial \mathcal{L}}{\partial e_{a \lambda}} =- \kappa \sqrt{-g}\: e^{a}_{\mu}  \Big ( 4 \Sigma^{bc \lambda} T_{bc}^{\; \; \mu} -g^{\lambda \mu} \Sigma^{bcd} T_{bcd} \Big ),
\end{equation}
being $\delta\mathcal{L}_M/ \delta e_{a\rho}:= \sqrt{-g} \: e^{a}_{\nu} \Theta^{\nu}_{\rho}$ the matter energy- momentum tensor and $t^{\lambda \mu}= t^{\lambda a} e_{a}^{\mu}$ the gravity energy-momentum tensor. Because of the local Lorentz invariance of the field equations, there is a six-fold degeneracy in the theory given by the six parameters of the local Lorentz transformation (rotations and boosts in tangent space). In other words, the theory only fix the metric structure. 

From the teleparallel field equations (\ref{eq: campos}), considering the asymmetry of the superpotential in the last two indices, we obtain the conservation law:
\begin{equation}
\partial_{\mu} \Big [ \sqrt{-g} \Big (t^{\mu a} + \Theta^{\mu a} \Big ) \Big ] = 0.
\label{eq: conservacion}
\end{equation}

This is a regular conservation law ---not a local covariant one--- from which we can define conserved charges associated with gravity plus matter:
\begin{equation}
P^{a} = \int_{\mathcal{D}} (t^{\lambda a} +\Theta^{\lambda a} ) d\mathcal{D}_{\lambda} \equiv 4 \kappa \oint_{\partial \mathcal{D}} \Sigma^{a \lambda \sigma} dS_{\lambda \sigma}.
\label{eq: tetraimpulso}
\end{equation}

Here, $\mathcal{D}$ is a 3-dimensional hypersurface and $\partial \mathcal{D}$ is its 2 dimensional boundary; we have used Stoke's theorem in the last step. The $a=(0)$ component, the projection of the energy-momentum onto the time-like component of the tetrad, is the total energy:
\begin{equation}
E= P^{(0)}=  \int_{\mathcal{D}} (t^{\lambda \mu} +\Theta^{\lambda \mu} ) \: e^{(0)}_{\mu} \: d\mathcal{D}_{\lambda}.
\end{equation}

The teleparallel gravitational energy has been much investigated in recent years, being a simpler and more straightforward approach than other geometric quasi-local treatments of the gravitational energy \cite{Maluf:2005kn}. The physical interpretation of the energy concept in GR$_{||}$ depends on the role of the tetrad in the theory (see Ref. \cite{Maluf:2007qq} and Ref. \cite{luca} for two different approaches). This is also a relevant issue for understanding the alternative theories of gravity that rely on the concept of absolute parallelism, such as $f(\mathbb{T})$ and non-local gravity. 

In the next section, we present an axiomatic formulation of GR$_{||}$. We will then use this axiomatization for comparing the teleparallel formalism with General Relativity and proving the physical inequivalence between both theories.

\section{Axiomatics}

Besides a formal set up, a physical theory is endowed with semantic	assumptions that establish its physical content. These assumptions, however, are usually presented in heuristic manner. If we wish to investigate the key concepts of a theory, the axiomatic format is convenient (see Ref. \cite{bunge2013foundations}). The dual axiomatic method exactifies and systematizes the physical and formal content in an axiomatic basis \cite{Bunge2016}. The axiomatic basis can be written as
\begin{equation*}
	\mathcal{A}=\mathcal{A}_F \wedge \mathcal{A}_S \wedge \mathcal{A}_P,
\end{equation*}
where $\mathcal{A}_F$ are the formal axioms---of purely mathematical content--- $\mathcal{A}_S$ are the semantic axioms---relating mathematical constructs with factual objects--- and $\mathcal{A}_P$ are the physical axioms--- expressing relations among constructs that represent physical entities, e.g. the dynamical equations of the theory. All theories also assume a formal and ontological background on which they are based.

A physical theory $T$ is then a set of statements $s$ closed under logic implication from the axiomatic basis $\mathcal{A}$, i.e,
\begin{equation}
T= \Big \lbrace s \: : \: \mathcal{A} \rightarrow s \Big \rbrace.
\end{equation}

The referents of the theory should be explicitly stated in the axioms. In this sense, we can build different theories, or interpretations of a theory, holding the formal apparatus and changing the semantic postulates.

In the case of GR$_{||}$, a different formalism from GR is used. Since the dynamical equations are equivalent to Einstein's equations, it is assumed that both theories are equivalent. However, if the alternative formalism introduces new semantical axioms, novel aspects of the physical system may appear that will only be represented in the new theory. A strict equivalence between two theories would hold iff all axioms in each theory are obtained from the other, i.e. $\mathcal{A}(GR_{||}) \longleftrightarrow \mathcal{A}(GR)$. The purpose of the next sections is to verify whether this is true for GR and GR$_{||}$. Also, we will use our axiomatization to show the scope and relevance of the tetrad formalism in GR$_{||}$ and other teleparallel theories.

\subsection{GR$_{||}$ axiomatized}

In order to formulate a realistic axiomatization of GR$_{||}$, we will first make explicit the background concepts. Then we will establish the generating basis from which we will construct the axiomatic formulation of the theory.

\subsection*{Background}

The formal background of the theory includes first order logic, mathematical analysis, and differential geometry. Let us note that space-time theories like GR are more fundamental than other field theories since all matter fields 'live' onto space-time, i.e. most physical theories assume a background space-time in its foundations. In this sense, a dynamical theory of space-time is more closely related to an ontological theory \cite{Romero2015}.

\subsection*{Generating basis and definitions}

The generating basis of GR$_{||}$ is constructed with 10 elements

\begin{equation}
\mathcal{B}= \lbrace \mathcal{ST}, \Sigma, \mathbb{K} , 	\mathcal{M}, \lbrace \mathbf{g} \rbrace , \lbrace \textbf{e}_a \rbrace, \lbrace \Theta \rbrace, \lbrace \phi \rbrace , \lbrace \Lambda^{a'}_{b}\rbrace,\kappa\rbrace.
\end{equation}

The meaning of these symbols will be given in the axiomatic basis. First, we set out the main definitions of Weitzenb\"ock tensors used in the theory:

\begin{description}[0.5cm]
	\item[D$_{1}$] $T^{\rho}_{\mu \nu}:= e^{\rho}_a \Big ( \partial_{\nu} e^a_{\mu} - \partial_{\mu} e^a_{\nu} \Big )$ is the torsion tensor
	\item[D$_{2}$] $K_{\; \mu \nu}^{\rho}:= \frac{1}{2} g^{\rho \sigma} \Big ( T_{\nu \rho \mu} + T_{\mu \sigma \nu} - T_{\sigma \mu \nu} \Big )$ is the contorsion tensor.
	\item[D$_{3}$] $\Sigma^{\rho \mu \nu}:= \frac{1}{2} \Big ( K^{\mu \nu \rho} -g^{\rho \nu} T^{\sigma \mu}_{\; \; \; \sigma} + g^{\rho \mu} T^{\sigma \nu}_{\; \; \; \sigma} \Big )$ is the  \textit{superpotencial}.
	\item[D$_{4}$] $t^{\lambda \mu} := \kappa \Big ( 4\Sigma^{bc \lambda} T_{bc}^{\; \; \; \mu} -g^{\lambda \mu} \Sigma^{bcd} T_{bcd} \Big )$ is the energy-momentum tensor of space-time.
\end{description} 
\subsection*{Axiomatic Basis}
The axiomatic basis is given by $\mathcal{A}($GR$_{||}) = \bigwedge \limits_{i}^{16} \mathbf{A}_i$. We present the axioms divided in four groups. The axioms are formal (FA), semantical (SA), or physical (PA)\footnote{For other theories presented in a similar way see Refs. \cite{PerezBergliaffa:1993kz}, \cite{PerezBergliaffa:1995ky}, and \cite{Covarrubias1993}.}

\subsection*{Axioms}
\textit{Group I: Space-time}

\begin{description}[0.5cm]
	\item[A$_1$] (FA) $\mathcal{M}$ is a Hausdorff para-compact, $C^{\infty}$, 4-dimensional, real and pseudo-Riemannian manifold.
	\item[A$_2$] (FA) $\lbrace \textbf{g} \rbrace $ is a family of rank-2 metric tensors, symmetric, and $+2$ signature. All minor principals of the metric tensor $g^{\mu \nu}$ are negative.
	\item[A$_3$] (FA) $\lbrace \phi \rbrace $ is a family of isometries: $\phi^{*} \textbf{g}=\textbf{g}$.
	\item[A$_4$] (SA) Space-time $\mathcal{ST}$ is the physical system represented by the equivalence class of isometric diffeomorphism of a given metric, i.e. $\mathcal{ST}$$\widehat{=}$ $\overline{( \mathcal{M}, \textbf{g} )}$ \footnote{The symbol $\hat{=}$ is used for the relation of representation. See Ref. \cite{bunge1973philosophy} for details.}
\end{description}

\bigskip 

\textit{Group II: Matter}

\begin{description}[0.5cm]
	\item[A$_{5}$] (FA) $\Sigma$ is a non-empty set of objects $\sigma \in \Sigma$.
	\item[A$_{6}$] (SA) There is an element $\square \in \Sigma$ which denote the absence of physical system. For all $\sigma \in \Sigma$ other than $\square$, $\sigma$  denotes a physical system different from space-time.
	\item[A$_{7}$] (FA) For each $\sigma \in \Sigma$ there is a symmetric 2-rank tensor field $\Theta$. In particular, there is a one-to-one correspondance between $\square \in \Sigma$ and the null tensor field $\Theta=0$
	\item[A$_{8}$] (SA) $\Theta$ represents the energy-momentum tensor of the physical system $\sigma$.
\end{description}

\bigskip 

\textit{Gruop III: Reference system}
\begin{description}[0.5cm]
	\item[A$_{9}$] (FA) The tetrad $\lbrace \textbf{e}_a(x) \rbrace$ is an orthonormal basis of $T_p \mathcal{M}$ in each point of the manifold: $e^{a}_{\mu} (x) e^{x}_{\nu}(x) \eta_{ab} = g_{\mu \nu}(x)$.
	\item[A$_{10}$] (FA) $\mathbb{K} \subset \Sigma$ is a non empty family of objects $K \in \mathbb{K}$. We call \textit{frames} the elements of $\mathbb{K}$.
	\item[A$_{11}$] (SA) A reference frame $K$ is represented by a time-like congruence $\mathcal{C}$ and a tetrad field $\lbrace \textbf{e}_a(p), p\in \mathcal{C} \rbrace$  where $\bold{e}^{q}_{\; (0)} \equiv \bold{U}^{q}$ is the tangent vector of the curve $\gamma_{q} \subset \mathcal{C}$, i.e $\langle \mathcal{C}, \bold{e}_a \rangle$ $\widehat{=}$ $K$. 
	\item[A$_{12}$] (FA) $\lbrace \Lambda^{b'}_{a}(x) \rbrace$ is a family of point-dependent Lorentz transformations.
	\item[A$_{13}$] (SA) $\forall \: K, K^{'} \in \mathbb{K}, \exists \: \Lambda^{b'}_{a}(x)$ such that if $\langle \mathcal{C}', \bold{e}_{a'} \rangle $$\widehat{=}$$K'$ and $\langle \mathcal{C}, \bold{e}_a \rangle$ $\widehat{=}$ $K$ then $\Lambda^{b'}_{a}(x)\: \textbf{e}_{b'}=\textbf{e}_{a}$, $\forall p \in \mathcal{C} \cap \mathcal{C'}$.
	\item[A$_{14}$] (SA) Let $x$ a physical system and $\mathcal{P}_{x}$ a given property of $x$. If $\mathcal{P}_{x}$ is represented with a tensor field $\mathbf{P} \hat{=} \mathcal{P}_x$, the values of this property in a reference frame $K$ are obtained from the projected tensor on the tetrad field, $ \mathbf{P}\cdot \mathbf{e} \: \hat{=} \: \mathcal{P}_x (K)$.  
\end{description}

\bigskip 

\textit{Group IV: Dynamics}

\begin{description}[0.5cm]
	\item[A$_{15}$] (SA) $\kappa \in \mathbb{R}$, where $[\kappa]= L \: M \: T^{2}$.
	\item[A$_{16}$] (PA) A reference frame $K$ is constrained by the Einstein Teleparallel equations:
	\begin{equation*}
	\partial_{\nu} \Big ( \sqrt{-g} \: \Sigma^{a \lambda \nu} \Big ) = \frac{\sqrt{-g}}{4k} \: e^{a}_{\mu} \Big ( t^{\lambda \mu} + \Theta^{\lambda \mu}  \Big ).
	\end{equation*}
\end{description}

\section{Discussion}

The axiomatic system constructed above for GR$_{||}$ will allow us to analyze three relevant topics: (i) the physical meaning of the teleparallel structure of the theory and its equivalence to GR, (ii) the role of the tetrad field in characterizing a local energy-momentum tensor for gravity, and finally (iii) the main differences between GR$_{||}$ and other teleparallel theories.  We begin with the referents of the theory, i.e. what kind of physical systems the theory describes.

\subsection{Reference class}

The reference class of an axiomatized theory is closed \cite{BSem}. In the case of GR$_{||}$, the class is formed by space-time and ordinary matter:
\begin{equation}
\mathcal{R} \Big (\bigwedge \limits_{i}^{15} \mathbf{A}_i \Big ) = \bigwedge\limits_{i}^{15}  \mathcal{R}  \Big (\mathbf{A}_i \Big) = \lbrace \Sigma, \mathcal{ST} \rbrace.
\end{equation}

We stress that in our axiomatic formulation, space-time is a physical entity endowed with properties. These properties are associated with the topological character of the manifold (e.g. compactness), with the metric, or with a mix of both (e.g. time-orientability). Space-time is connected with the dynamical object of GR$_{||}$ in two ways: the tetrad field is defined over the manifold $\mathcal{M}$ and is related to the metric by the orthonormal relation ($\mathbf{A}_{8}$). The latter implies that the dynamical equations of the theory impose restrictions to the metric properties of space-time, and these restrictions are equivalent to Einstein's equations. A more elegant way to show the association among frames, the Lorentz group, and the manifold, is through the fibre bundle framework of space-time, a path we will not follow in this work \cite{prugovecki2013quantum}.

Matter enters in the theory represented by the energy-momentum tensor and also in the characterization of the reference frame ($\mathbf{A}_{8}$ and $\mathbf{A}_{10}$). The dynamics of fields and particles must be supplemented by other theories, while GR$_{||}$ (and GR) is only concerned with their energy-momentum distribution.\\

\textbf{Remark 1} A reference system is not another \textit{kind} of matter in our axiomatization. Group III of axioms only implies that local properties of physical systems are relative to other physical systems. We followed the approach where the reference frame is represented by an orthonormal basis field and not by a coordinate system \cite{Felice:2010cra}. What kind of physical objects are convenient to be adopted as a reference frame is a question that requires further analysis. \\

From axiomatization of GR (see $\mathcal{A}($GR$)$ in Appendix A), we immediately verify that both GR and GR$_{||}$ share the same referents. It is then appropriate a thorough comparison of both theories since they have identical domain.

\subsection{Recovering General Relativity}

While the metric field characterizes space-time, the tetrad field establishes how local properties behave. Both are linked by
\begin{equation}
g_{\mu \nu}=e^{a}_{\mu}e^{b}_{\nu} \eta_{ab}.
\end{equation}
which means that a given space-time allows infinite reference frames related by local Lorentz transformation. Conversely to all other teleparallel theories, GR$_{||}$ equations are local Lorentz invariant, equivalent to Einstein's equations.\\

\textbf{Theorem 1} GR$_{||}$ equations for $\mathbf{e}^{a}$ are equivalent to GR equations for $g_{\mu \nu}$.\\

An sketch of the proof is given in Ref.\cite{local} and also in Ref.\cite{Maluf:2013gaa}. The equivalence of (\ref{eq: campos}) to Einstein's equations gives the following corollary\\

\textbf{Corollary 1} GR$_{||}$ equations are diffeomorfic and local Lorentz invariant.\\

These two types of invariance have very different physical meanings. \\

$\diamond$ Diffeomorfism invariance is associated with the nature of space-time. Conversely to Newtonian theories, space-time should not be regarded as an absolute stage where other physical fields live. Instead, in GR, space-time is the equivalence class of the pair $ ( \mathcal{M}, g_{\mu \nu} )$ under active diffeomorphisms. One should be careful to assign physical meaning to points $p \in \mathcal{M}$ or coordinates, since only their relation to $g_{\mu \nu}$ is meaningful\footnote{As Rovelli states in Ref. \cite{rovelli2007quantum}, in the general relativistic point of view we only have 'fields over fields'. However, it seems to us that space-time has very distinctive features from other matter fields. Because of this, we have chosen the term 'space-time' and not 'gravitational field' to denote $\mathcal{ST}$. We say then 'fields over space-time'}. As a consequence, the field equations of the theory are coordinate-invariant.\\

$\diamond$ Local Lorentz invariance (LLI) becomes apparent only after we introduce the tetrad field in our theory. Taking into account axioms $\mathbf{A}_{13}$,$\mathbf{A}_{14}$, and Corollary 1, LLI means that space-time fixes all possible physical frames, i.e. all physical observers 'perceives' the same underlying space-time. Axioms $\mathbf{A}_{13}$ and $\mathbf{A}_{14}$ and their consistency with the field equations via LLI is known as the \textit{Hypothesis of Locality}. In words of Mashhoon, this hypothesis states that 'an accelerated observer is pointwise inertial', no matter the history of the observer \cite{Mashhoon:2012az}. If we consider a space-time theory with no LLI field equations for $\mathbf{e}^{a}$, then either Group I or Group III of axioms must be revised. We will return to this point in the last section.\\

We have shown that a tetrad field associated with a reference system that is solution of eqs. (\ref{eq: campos}) fixes the metric and therefore space-time. Given the equivalence of $\mathbf{A}_{16}$ to Einstein's equations, all dynamical features of GR might be derived from GR$_{||}$ \\

\textbf{Theorem 2} GR$_{||}$ implies GR;  $\mathcal{A}(GR_{||}) \Rightarrow \mathcal{A}(GR)$.\\

Now, we move on to analyze the new features that the tetrad field introduces in GR$_{||}$ and their relation to the teleparallel geometry.

\subsection{Change and geometry}

The concept of change is intimately associated with the nature of space-time. In order to formulate general relativistic laws describing changes in physical properties we must take into account that (i) the laws must be consistent with the adopted space-time representation and the hypothesis of locality and (ii) these laws should characterize changes \textit{over} space-time. The first requirement states that the laws of physics should be diffeomorfic and local Lorentz invariant. The second condition, related to the first one, imposes that these changes ought to be formulated in a metric way (space-time is represented solely by the metric and the manifold in our axiomatization (Group I)); a natural procedure to do this is introducing a metric covariant derivative $\nabla$, i.e. the Levi-Civita connection $\Gamma$. Other non-metrical derivatives could introduce additional degrees of freedom (e.g. Einstein-Cartan connections) or lack sufficient structure (e.g. Lie derivatives). 

A reference frame system may be then analyzed by its covariant derivative $\nabla_{\nu}e^{a}_{\mu}$, which characterize how it behaves with respect to space-time. Let us consider the directional derivatives of $\mathbf{e}^{a}$ in the direction of another tetrad $\mathbf{e}^{c}$, which can be written as:
\begin{equation}
e^{\nu}_c \nabla_{\nu} e^{a}_{\mu} = K^{a}_{bc} e^{c}_{\mu},
\label{eq: contortionlevi}
\end{equation}
where $K^{a}_{bc}$ are the projected components of the contortion tensor \cite{Sotiriou:2010mv}. Thus, the contortion tensor can be identified with the Ricci rotation coefficients, usually used in the $1+3$ orthonormal frame approach to characterize the kinematical properties of the tetrad field \cite{Felice:2010cra}. The change of the tetrad over its time-like velocity component $\mathbf{e}^{(0)} \equiv \mathbf{U}$ is
\begin{equation}
U^{\nu} \nabla_{\nu} e^{a}_{\mu} = K^{a}_{b (0)} e^{b}_{\mu} = \phi^{a}_{b} e^{b}_{\mu},
\label{eq: acceleration}
\end{equation}
being $\phi_{ab}:=K_{ab(0)}$ the antisymmetric acceleration tensor. In analogy to Faraday's electromagnetic tensor, we can decomposed $\phi_{ab}$ in a translational part $a_{(i)}:=\phi_{(0)(i)}$ and a rotational part $\phi_{(i)(j)} = \epsilon_{ijk} \Omega^{k}$. If we take the dynamics of a free particle, say, from a variational principle, it is well known that it would follow the Riemannian geodesic equation
\begin{equation}
U^{\nu} \nabla_{\nu} U_{\mu}=0.
\end{equation}

This means that if the translational part of the acceleration $a_{(i)}$ tensor vanishes, the reference frame is in free fall: the congruence is composed of geodesics of the Riemann geometry. Furthermore, if the rotational part is null, the reference frame system is not rotating (with respect to a Fermi-Walker transported frame). Thus we adopt the following definition\\

\textbf{Definition 1} If the acceleration tensor $\phi_{ab}$ is null over the congruence $\mathcal{C}$, the reference system frame is a \textit{pseudo inertial reference frame} (PIRF).\\

\textbf{Remark 2} The acceleration tensor is not covariant under local Lorentz transformations, as it is easily seen from their relation to the non-covariant torsion tensor \cite{Maluf:2007qq}. Hence, when we perform a general local Lorentz transformation onto a frame, all inertial properties change unless the Lorentz transformation is global. Nevertheless, the relation of a physical quantity from any reference system to another, e.g. from an inertial one to one that is accelerated, is strictly local, consistent with the hypothesis of locality.\\

The Weitzenb\"ock connection $^{*} \Gamma$, on the other hand, defines a covariant derivative that quantifies how a given tensor $\mathbf{P}$ changes with respect to a preferred tetrad frame. If this quantity $\mathbf{P}$ is fixed on the tetrad, (i.e. if $\mathbf{P} \cdot \mathbf{e}^{a}$ is constant) then its Weitzenb\"ock derivative is zero no matter the path chosen. This connection also defines an alternative concept of acceleration; for instance, a free falling path is, in general, accelerated in the Weitzenb\"ock geometry (see equation (\ref{eq: freefall}) below). The teleparallel force equation for a free falling particle with velocity $U^{\mu}$ is
\begin{equation}
U^{\nu} \: ^{*} \nabla_{\nu} U^{\mu} = \mathcal{F}_{T}^{\mu}, \quad \mathcal{F}_{T}^{\mu}:= K^{\mu}_{\nu \rho} U^{\nu} U^{\rho}.
\label{eq: freefall}
\end{equation}

Note that the term $\mathcal{F}_{T}$ is a pseudo-force, i.e. frame-dependent. For instance, if we choose a free falling frame comoving with the particle, from (\ref{eq: acceleration}) we obtain $U^{\nu} \: ^{*} \nabla_{\nu} U^{\mu}=0$; this means that the weak equivalence principle is still satisfied in GR$_{||}$. On the other hand, real interaction terms, such as the Lorentz force $\mathcal{F}_{L}:=F^{\mu \nu} U_{\mu}$, are always frame-independent and imprint an absolute (Levi-Civita's) acceleration to the system. 

At this point, it is useful to introduce the following definition\\

\textbf{Definition 2} A tensor field $\mathbf{P}$ associated with a physical system $S$ is an \textit{intrinsic} property of $S$ iff $\mathbf{P}[\mathbf{e}^{a}]=\mathbf{P}[\mathbf{e}^{b'}]$, for any tetrad field.\\

Note that even if $\mathbf{P}$ is an intrinsic property, we can obtain the values of this property in a given frame by projecting the tensor field onto the tetrads ($\mathbf{A}_{14}$). However, the relation between frames of an intrinsic property is covariant and local. This is not the case of the Weitzenb\"ock torsion tensor, which is frame-dependent. The Riemann curvature tensor, on the contrary, is an intrinsic property of space-time. For instance, if the curvature tensor is zero in one frame, it is zero for all observers. Torsion, however, might be non zero even in Minkwoski space-time, if we choose an anholonomous tetrad, e.g. one representing an accelerating observer. We conclude that torsion is not a proper quantity to represent the gravitational interaction in GR$_{||}$; it characterizes kinematic properties of reference frames (see Eq. \ref{eq: contortion}) and it is constrained over a congruence by the field equations (29), i.e. constrained by the underlying space-time.

\subsection{Locality of the energy-momentum tensor for gravity}

We have seen in the previous sections how the tetrad field and the teleparallel geometry of GR$_{||}$ encode a broader representation than the metric field. Let us return now to the dynamics of the theory. The essential feature of GR$_{||}$ field equations is the possibility of deriving an energy-momentum conservation law for matter$+$gravity that is not attainable in GR:\\

\textbf{Theorem 3} Giving a tetrad field $e^{\mu}_{a}$, the teleparallel equations admit a conservation equation for each $a$ component:
\begin{equation*}
	\partial_{\lambda} \partial_{\nu} \Big ( \sqrt{-g} \: \Sigma^{a \lambda \nu} \Big ) =0 \rightarrow \nabla_{\lambda} ( t^{\lambda a} + \Theta^{\lambda a}  )=0.
	\label{eq: campo}
\end{equation*}

In this equation, we associate $t^{\lambda a}$ to the energy-momentum tensor of space-time, consistent with its Lagrangian definition (23). This teleparallel energy-momentum tensor has been studied by several authors (see Ref. \cite{deAndrade:2000kr}). Our intention here is to characterize $t^{\mu \nu}$ under our axiomatic formulation. The usual approach to define pseudotensors in GR is similar to the teleparallel case: the Ricci scalar is decomposed in a linear-derivative part $\tilde{\mathcal{L}}$ and a total divergence,
\begin{equation*}
	R = \tilde{\mathcal{L}} + \nabla_{\mu} \tilde{\mathcal{W}}^{\mu},
	\label{eq: sep}
\end{equation*}
to obtain a superpotential equation similar to (\ref{eq: campos}):
\begin{equation}
	\partial_{\sigma} ( \sqrt{-g} \:  S^{\mu \rho \sigma} ) = \sqrt{-g} \: ( \tau ^{\mu \rho} + \Theta^{\mu \rho} ).
	\label{eq: pseudoec}
\end{equation}

The main difference between these approaches and GR$_{||}$ is that $t^{\lambda a}$ is a well-behaved tensor under coordinate transformations. This is true because GR$_{||}$ has a first-order diffeomorphic-invariant Lagrangian. In this sense, GR$_{||}$ is a better theory to formulate an energy-momentum tensor. Even though the teleparallel energy definition (26) is consistent in many physical scenarios, the local behavior of the teleparallel energy-momentum tensor remains unclear for two reasons: (i) the gauge freedom for choosing a tetrad field (non-covariance under LLT) and (ii) its compatibility with the equivalence principle, that is, that gravity does not manifest locally in a free-falling frame, for which local approaches to define an energy-momentum density are usually discarded.

In contrast to the matter energy-momentum $\Theta^{\mu \nu}$, the space-time energy-momentum tensor $t^{\mu a}$ is not an intrinsic property, in the sense of Definition 2. This feature is expected, because energy and momentum are frame-dependent properties and frame properties are ultimately determined by space-time itself. The lack of a preferred reference frame in curved space-time makes the energy-momentum tensor extremely degenerated as we will show below. Thus, in the context of GR$_{||}$, to characterize the tensor\footnote{Note that $t^{\mu \nu}$ is not the pseudotensor obtained writing the field equation as equation (\ref{eq: pseudoec}). In the notation used in Ref\cite{deAndrade:2000kr}, our $t^{\mu \nu}$ corresponds the contravariant version of $h_{a}^{\mu} j^{a}_{\nu}$} $t^{\mu \nu}$ we must first completely characterize the frame (i.e. the tetrad field, see Group III).

It is often stated in the literature that a frame can be fixed if its kinematical properties encoded in the acceleration tensor $\phi_{ab}$ are fixed \cite{Maluf:2013gaa}. However, this is not entirely correct since the full kinematics of a frame is contained in $K^{a}_{bc}$ rather than $\phi_{ab}$. For example, the contortion tensor contain the expansion $\theta_{ab}$ and vorticity $\omega_{ab}$ of the congruence,
\begin{equation}
\omega_{ab}:= K_{(0)(ab)}, \quad \theta_{ab}:=K_{(0)[ab]},
\end{equation}
as well as the spatial motion of the frame $K_{(i)(j)(k)}$. In other words, we should take into account the whole congruence $\mathcal{C}$ and not just a path in order to describe properties of extended objects as fields. In order to illustrate the importance of this observation in the teleparallel framework, it would be useful to introduce the distinction made in Ref. \cite{Giglio:2011gk} between Local Reference Frame and the Pseudo Inertial Reference Frame of Def. 1:\\

\textbf{Definition 3} Given a congruence $\mathcal{C}$ with an associated tetrad field $\mathbf{e}^{a}$ and a timelike geodesic $\gamma \subset \mathcal{C}$, the reference frame $K$ is a local inertial reference frame (LIRF), associated with $\gamma$ if
\begin{equation}
\mathbf{e}_{a}\mid_{\gamma} = \partial_{\hat{\mu}}\mid_{\gamma},
\end{equation}
where $\lbrace x^{\hat{\mu}} \rbrace$ are Fermi coordinates,
\begin{equation}
g_{\hat{\mu}\hat{\nu}} \mid_{\gamma}=\eta_{\hat{\mu}\hat{\nu}}\mid_{\gamma}, \quad \partial_{\hat{\rho}} g_{\hat{\mu}\hat{\nu}}\mid_{\gamma}=0.
\end{equation}

Thus, the congruence of a LIRF has zero torsion $T^{\hat{\mu}}_{\hat{\nu} \hat{\rho}}\mid_{\gamma}=0$ because the frame is holonomous over the geodesic (this is true in all coordinate system). Outside this path however, torsion is non-zero. We can prove in addition that this LIRF is composed of accelerated curves outside $\gamma$, whose acceleration tensor is related directly to the Riemann curvature \cite{Bini:2015xqa}. Even though a LIRF and a PIRF can contain a same geodesic observer moving over a path $\gamma$, we obtain in general:
\begin{equation}
t^{\mu \nu}_{\text{PIRF}} \mid_{\gamma}  =0, \quad t^{\mu \nu}_{\text{LIRF}}\mid_{\gamma} \neq 0.
\end{equation}

This fact shows that even though the energy-momentum tensors are evaluated on the same non-rotating geodesic path, they depend on the kinematic properties of the whole congruence encoded in the contorsion tensor, and not only in its acceleration. We illustrate this in a concrete example on a Schwarzchild space-time in Appendix B. 

A LIRF is a special kind of frame where all accelerations and kinematical properties of the congruence are related only to the curvature tensor. This frame seems to behave appropriately since (i) it is zero on a given geodesic, according to the equivalence principle, and (ii) is related to the Bel-Robinson tensor at second order (see Ref. \cite{So:2008kr}). Even though we have used a LIRF to show and compare the local properties of the energy-momentum, this frame is only useful to integrate physical quantities over a small region around the geodesic. If we want to obtain the total energy of space-time over a time slice, we need to use a general frame such as a PIRF. In these type of frames there are inertial contributions to the total energy. These contributions are represented in the contortion tensor, which is ultimately constrained by the underlying space-time---in our example in Ap. B, the frame $K'$ has non-null shear components. In short, if we want to study the energy of space-time measured by an observer in some state of inertia, we still have infinite ways to choose this observer in an extended region. This sort of degeneracy of $t^{\mu \nu}$ disappears if the theory fixes completely the tetrad field or if a physical criterion for choosing the frame is founded.

\subsection{Inequivalence between GR and GR$_{||}$}

Now, we summarize the two main points made so far in this work:\\

$\diamond$ GR$_{||}$ has a dynamical object that encodes a broader representation than the metric, i.e. the tetrad field representing a reference frame.\\

$\diamond$ GR$_{||}$ dynamical equations allow to derive an energy-momentum conservation law for matter plus gravity. The gauge freedom in the election of the energy-momentum tensor for gravity and its non covariant character can be interpreted as the free choice of a reference frame system. \\

In General Relativity, reference frames are absent from the core of theory and, for this reason, a well-defined conservation of energy-momentum cannot be provided. We conclude:\\

\textbf{Theorem 4} Not all the axioms of GR$_{||}$ can be derived from the axioms of General Relativity, $\mathcal{A}(GR) \nRightarrow \mathcal{A}(GR_{||})$. Therefore, the theories are physically inequivalent.\\

GR and GR$_{||}$ are both theories of space-time and matter. GR is only concerned with intrinsic properties of space-time, whereas GR$_{||}$ includes the notion of the reference frame in the basis of the theory. 

Our discussion was centered around the metric formulation of GR. However, there are several fundamental tetrad formulations of GR (that is, with Riemannian geometry and Einstein's equations), for instance, those of Newman-Penrose (NP), Geroch-Held-Penrose (GHP), and the so-called 1+3 orthonormal approach. A straightforward reconstruction of GR adopting one of these formulations is not directly equivalent to the axiomatic scheme presented in Sect. 3 because in most cases the physical interpretation of the tetrads is different. For instance, in the NP and in the GHP formalisms null tetrads are used, and hence they cannot be interpreted as reference frames as we did through Group 3 of axioms. Even though the Einstein's equations are included in the dynamics of these theories, in all these cases there are more degrees of freedom than ordinary metric GR. The key point for our analysis here is whether the new formalism introduces new physical features in the basis of the theory.

In the case of GR$_{||}$, the redundancy of the tetrad field is used to interpret these dynamical objects as reference frames. Together with the teleparallel field equations, this interpretation is used to account for the conservation of the gravitational energy-momentum tensor, that holds for each frame (in a non covariant way). On the other hand, in the 1+3 approach, the reference frame interpretation of tetrads may hold, but the geometrical set up differs from GR$_{||}$. In this way, the construction of a gravitational energy-momentum tensor of this sort cannot be achieved in these formulations.

\section{On other teleparallel theories}

In recent years there has been an increasing interest in teleparallel theories. Two of the theories that have atracted attention are the so-called $f(\mathbb{T})$ \cite{Ferraro:2011us} and non-local gravity \cite{Mashhoon:2014twa}. 

In $f(\mathbb{T})$ theories, the teleparallel Lagrangian of GR$_{||}$ is modified analogously to $f(R)$ theories. The resulting teleparallel equations are of second order, contrary to the fourth order equations of $f(R)$
\begin{equation}
\partial_{\sigma} \Big ( \sqrt{-g} \Sigma^{a \lambda \sigma} f'(\mathbb{T})  \Big) =\frac{\sqrt{-g}}{4\kappa} \Big( \mathcal{T}^{\lambda a} + \Theta^{\lambda a} \Big ),
\end{equation}
where $\mathcal{T}^{\lambda \mu}:= \kappa \Big ( 4 f'(\mathbb{T}) \Sigma^{b c \lambda} T_{bc}^{\; \; \mu} - g^{\lambda \mu} f(\mathbb{T}) \Big)$ is the modified energy-momentum tensor. This theory has been widely applied to cosmology, where the observed acceleration of the universe is explained without introducing dark energy. Contrary to GR$_{||}$, the theory lacks local Lorentz invariance\cite{Sotiriou:2010mv}, which means that different tetrads, related by local Lorentz transformations, have different motion equations. In this sense, $f(\mathbb{T})$ has more degrees of freedom than GR$_{||}$. 

It is an interesting question whether the tetrad field in this theory can represent a reference frame in the same way as it was constructed in our axiomatization of GR$_{||}$. The main difference with GR$_{||}$ is that each kind of observer would 'sense' a different underlying space-time since each tetrad field fixes a different metric. The transformation laws between frames as well as the representation of space-time should be modified to encompass this interpretation. In the current state of the theory, the tetrad field is only used to obtain solutions of the field equations from which a metric field is obtained. The notion of the tetrad field as denotating a reference frame has not been studied yet.

On the other hand, non-local gravity is more connected with the GR$_{||}$ approach\cite{Mashhoon:2014twa}. The axioms of GR$_{||}$ imply that frames are related in a local way, meaning that an accelerated observer is pointwise inertial. Mashhoon has argued (following Bohr and Rosenfeld \cite{Mashhoon:2003st}) that in classical field theories, a property of a given field, e.g. the Faraday tensor $F^{\mu \nu}$, cannot be measured in a pointwise manner and an averaging procedure over the history of the observer is needed. Hence the laws of physics must be rendered non-local (history-dependent).

Following what was made in non-local electrodynamics, and using the close analogy between GR$_{||}$ and Maxwell equations, the non-local gravity equations are obtained changing the superpotential:
\begin{equation}
\Sigma^{\mu \nu \rho} \rightarrow \mathcal{H}^{\mu \nu \rho}:= \Sigma^{\mu \nu \rho} + N^{\mu \nu \rho},
\end{equation}
where $N^{\mu \nu \rho}$ is an antisymmetric tensor involving the past history of the field. As an interesting consequence of non-local gravity, it is possible to show that observational data associated with dark matter might be explained as a non local effect.

In the teleparallel framework for gravitational theories, it is not clear whether we should hold the metric representation of space-time $\mathcal{ST}$. Instead, it seems more appropriate a tetrad-only form given by $\mathcal{ST}_{\text{T}}= \overline{(\mathcal{M}, \mathbf{e}^{a})}$. Recently, it is was argued by Schucking \cite{schucking2015einstein} that this tetrad-only representation, together with the elements of teleparallel geometry, imply Einstein first equivalence principle: the equivalence between acceleration and gravitation. For Schucking, this is explicit in the relation between the contorsion (i.e. the Ricci coefficients) and torsion (which represents, in this view, the gravitational interaction)

However, in GR$_{||}$ a pure tetrad representation of space-time is inadequate. Indeed, a frame-independent (metric-Riemannian) structure is always present in the theory constraining the acceleration fields $K^{a}_{\; bc}$. Nevertheless, it is an interesting subject to explore to what degree we can mimic curvature choosing specific frames, say, in flat space-time, and what is the role of torsion. It is possible to show that the geodesic deviation equation in a general frame is written as:
\begin{equation}
\ddot{\xi}_{a}= (-R_{(0)a(0)b} + C[\mathbf{e}^{a}]_{ab}) \: \xi^{b}
\end{equation}
where $C[\mathbf{e}^{a}]_{ab}$ is zero for LIRF \cite{Felice:2010cra}. Thus, in a general frame over flat space-time we would still measure geodesic deviation for some specific frames.

\section{Conclusions}

Using our axiomatic formulation, we have shown that\\

$(1)$ GR$_{||}$ is not fully equivalent to GR. Whereas all features of GR can be obtained from GR$_{||}$, the opposite is not true. The tetrad formalism is used to introduce the reference frame in the foundations of GR$_{||}$, allowing a consistent definition of a gravitational energy-momentum tensor and its conservation that are not attainable in GR. This energy-momentum tensor depends on the kinematic properties of the reference frame encoded in the contorsion tensor.\\

$(2)$ Both theories adopt a metric representation of space-time. In GR, the same metric is the dynamical object of the theory; however, GR$_{||}$ this part is played by the tetrad field. The Levi-Civita connection is the right formal tool to quantify changes over space-time, while the Weitzenbo\"ok connection measures how a given tensor changes with respect to a preferred frame. This might not be true in other alternative space-time theories, where the role of the tetrad field is different. For example, it is not clear whether $f(\mathbb{T})$ gravity shares the same physical interpretation of the tetrad as in GR$_{||}$. Further developments in such direction are needed.\\

Our axiomatization allowed us to explore some of the key physical concepts of GR$_{||}$ and compare them with GR and other teleparallel theories in a systematic way. Summing up, we found that is possible to construct two self-consistent theories describing the same physical entities, with equivalent dynamics, but with a different representation power.

\acknowledgments

This work was supported by grants PICT 2012-00878 (Agencia Nacional de Promoci\'on Cient\'ifica y Tecnol\'ogica, Argentina) and AYA 2013-47447-C3-1-P (Ministro de Educaci\'on, Cultura y Deporte, Espa\'na). We would like to thank Federico Lopez Armengol for helpful comments. We also thank the anonymous referees for suggestions and comments that helped improving the manuscript.

\appendix

\section{Axiomatization of General Relativity}

Here, we provide an axiomatization of GR to compare with the axiomatic formulation of GR$_{||}$ in Section 3. 

\subsection*{Generating basis and definitions}

The generating basis of GR is constructed with fewer elements than GR$_{||}$:

\begin{equation}
\mathcal{B'}= \lbrace \mathcal{ST}, \Sigma,\mathcal{M}, \lbrace \mathbf{g} \rbrace , \lbrace \Theta \rbrace, \lbrace \phi \rbrace , \kappa \rbrace.
\end{equation}

The meaning of these symbols will be given in the axiomatic basis. First, we set out the main definitions of Weitzenb\"ock tensors used in the theory:

\begin{description}[0.5cm]
	\item[D'$_{1}$] $R^{\rho}_{\mu \lambda \nu}:= \partial_{\lambda} \Gamma^{\rho}_{\mu \nu}- \partial_{\nu} \Gamma^{\rho}_{\mu \lambda} + \Gamma^{\rho}_{\sigma \lambda} \Gamma^{\sigma}_{\mu \nu}- \Gamma^{\rho}_{\sigma \nu} \Gamma^{\sigma}_{\mu \lambda}$ is the Riemann tensor.
	\item[D'$_{2}$] $R_{\mu \nu}:= R^{\rho}_{\mu \rho \nu}$ is the Ricci tensor.
	\item[D'$_{3}$] $R:= R^{\mu}_{\mu}$ is the Ricci scalar.
	\item[D'$_{4}$] $G_{\mu \nu}:= R_{\mu \nu} - \frac{1}{2} g_{\mu \nu} R$ is the Einstein tensor.
\end{description} 

\subsection*{Axiomatic Basis}
The axiomatic basis is given by $\mathcal{A}($GR$) = \bigwedge \limits_{i}^{10} \mathbf{A}_i$. We present the axioms divided in three groups. Note that Group I and II of axioms are exactly the same as in GR$_{||}$ but the reference system axiom group is absent. See Refs. \cite{Romero2015}, \cite{Covarrubias1993}, and \cite{bunge2013foundations} for similar axiomatizations of GR.
\subsection*{Axioms}

\textit{Group I: Space-time}

\begin{description}[0.5cm]
	\item[A$_1$] (FA) $\mathcal{M}$ is a Hausdorff para-compact, $C^{\infty}$, 4-dimensional, real and pseudo-Riemannian manifold.
	\item[A$_2$] (FA) $\lbrace \textbf{g} \rbrace $ is a family of rank-2 metric tensors, symmetric, and $+2$ signature. All minor principals of the metric tensor $g^{\mu \nu}$ are negative.
	\item[A$_3$] (FA) $\lbrace \phi \rbrace $ is a family of isometries: $\phi^{*} \textbf{g}=\textbf{g}$.
	\item[A$_4$] (SA) Space-time $\mathcal{ST}$ is the physical system represented by the equivalence class of isometric diffeomorphism of a given metric, i.e. $\mathcal{ST}$$\widehat{=}$ $\overline{( \mathcal{M}, \textbf{g} )}$.
\end{description}

\bigskip 

\textit{Grop II: Matter}

\begin{description}[0.5cm]
	\item[A$_{5}$] (FA) $\Sigma$ is a non-empty set of objects $\sigma \in \Sigma$.
	\item[A$_{6}$] (SA) There is an element $\square \in \Sigma$ which denote the absence of physical system. For all $\sigma \in \Sigma$ other than $\square$, $\sigma$  denotes a physical system different from space-time.
	\item[A$_{7}$] (FA) For each $\sigma \in \Sigma$ there is a symmetric 2-rank tensor field $\Theta$. In particular, there is a one-to-one correspondance between $\square \in \Sigma$ and the null tensor field $\Theta=0$
	\item[A$_{8}$] (SA) $\Theta$ represents the energy-momentum tensor of the physical system $\sigma$.
\end{description}

\bigskip 

\textit{Group III: Dynamics}

\begin{description}[0.5cm]
	\item[A$_{9}$] (SA) $\kappa \in \mathbb{R}$, where $[\kappa]= L \: M \: T^{2}$.
	\item[A$_{10}$] (PA) Space-time is constrained by the Einstein's equations:
	\begin{equation*}
	G^{\mu \nu}=\frac{1}{2\kappa} \Theta^{\mu \nu}.
	\end{equation*}
\end{description}

The direct inclusion of axiom Group III in GR$_{||}$ results in a redundancy given the Riemannian structure Einstein's equations (see \textbf{A}$_{10}$) and our previous analysis in Section 4. However, it might be necessary to include them if we want to analyze a quantum field theory in curved space-time. A group of axioms introducing tetrad field can be non-trivial if we change the axioms in Group III or the space-time representation. As was discussed before, some tetrad formulations are very useful to find exact solutions or deal with cosmological problems.

\section{Energy-momentum tensor for a LIRF and a PIRF}

Let us consider a Schwarzchild space-time and an arbitrary radial geodesic $\gamma_{*}$ on it. Over $\gamma_{*}$ we build two frames, a LIRF $K$, as in Definition 3, and a PIRF $K'$, which congruence $\mathcal{C}'$ is composed of radial non-rotating geodesic, $\phi_{a'b'}(p)=0$ for all $p\in\mathcal{M}$. We use an explicit realization of $K'$ presented in Ref. \cite{Maluf:2007qq} as:

\[ e^{\mu}_{(0)}= 
\left ( \begin{array}{c}
\alpha(r)\\
-\beta(r)  \\
0 \\
0  \end{array} \right), \quad e^{\mu}_{(1)}= 
\left ( \begin{array}{c}
\alpha(r) \beta(r) \cos(\phi)\sin(\theta)\\
\cos(\phi)\sin(\theta) \\
\cos(\theta) \cos(\phi)/r \\
- \text{csc} (\theta)	\sin(\phi)/r  \end{array} \right), \]

\[ e^{\mu}_{(2)}= 
\left ( \begin{array}{c}
\alpha(r) \beta(r) \sin(\theta)\sin(\phi)\\
\sin(\theta)\sin(\phi)  \\
\cos(\theta) \sin(\phi)/r \\
\cos(\phi) \text{csc}(\theta)/r  \end{array} \right), \quad e^{\mu}_{(3)}= 
\left ( \begin{array}{c}
\alpha(r) \beta(r) \cos(\theta) \\
\cos(\theta)  \\
-\sin(\theta)/r\\
0, \end{array} \right), \]
with $\alpha(r):=1/(1-2M/r)$ y $\beta(r):= \sqrt{2M/r}$. We can easily check that this free falling frame possess non-zero Weitzenb\"ock torsion. Its kinematical properties are encoded in the contortion tensor, whose non-null components represent the expansion of the congruence:

\begin{equation*}
K_{(0)(1)(1)} = \sqrt{\frac{M}{2r^3}}, \quad K_{(0)(2)(2)}=K_{(0)(3)(3)}= - \sqrt{\frac{2M}{r^3}}.
\end{equation*}

The teleparallel energy-momentum tensor can be calculated from definition $\mathbf{D}_{4}$. After some straightforward algebraic steps we obtain: 

\[ t_{\text{PIRF}}^{\mu \nu}=\left (
\begin{array}{cccc}
-\frac{M^2}{2 \pi  r^2 (-2 M+r)^2} & -\frac{\left(\frac{M}{r}\right)^{3/2}}{2 \sqrt{2} \pi  (2 M-r) r} & 0 & 0 \\
-\frac{\left(\frac{M}{r}\right)^{3/2}}{2 \sqrt{2} \pi  (2 M-r) r} & -\frac{M}{4 \pi  r^3} & 0 & 0 \\
0 & 0 & \frac{M}{8 \pi  r^5} & 0 \\
0 & 0 & 0 & \frac{M}{8 \pi  r^5} \\
\end{array} \right), \] 

The energy-momentum tensor is then symmetric, isotropic, and obviously non-zero through a radial geodesic $\gamma_{*} \subset \mathcal{C}'$. On the other hand, in a LIRF $K$, the energy-momentum tensor is zero on the radial geodesic $\gamma_{*}$
\begin{equation}
t^{\mu \nu}_{\text{LIRF}}\mid_{\gamma_{*}} =0
\label{eq: fermi}
\end{equation}
since we have zero torsion on $\gamma_{*}$ (note that equation (\ref{eq: fermi}) is valid in all coordinate system). We conclude that over the same radial geodesic path $\gamma_{*}$, the energy-momentum of gravity is zero in $K$ but non zero in $K'$. 

Since it is always possible to construct a LIRF, on any geodesic path, the field equations of the theory can be written as:
\begin{equation}
\partial_{\nu} \Big ( \sqrt{-g} \: \Sigma^{a \lambda \nu} \Big ) \mid_{\gamma} = \frac{\sqrt{-g}}{4k} \Big (\Theta^{\lambda a}  \Big )\mid_{\gamma},
\end{equation}
where $t^{\lambda \mu}\mid_{\gamma}=0$. This means that the matter energy-momentum is truly conserved over $\gamma$, 
\begin{equation}
\partial_{\lambda} \Theta^{\lambda a} \mid_{\gamma}=0
\end{equation}
and there is no interaction between gravity and matter in this frame. The equivalence principle in GR$_{||}$ can be stated as the following: There is always a reference system $K_{F}$ that contains a geodesic path $\gamma$ over which the energy-momentum of gravity vanishes.

\bibliographystyle{ieeetr}
\bibliography{bibliography}
% Please avoid comments such as "For a review'', "For some examples",
% "and references therein" or move them in the text. In general,
% please leave only references in the bibliography and move all
% accessory text in footnotes.

% Also, please have only one work for each \bibitem.

\end{document}